\RequirePackage{fix-cm} 
\documentclass[a4paper,twoside,11pt,dvips,reqno]{amsart}

\usepackage{fixltx2e}     
\usepackage{a4wide,indentfirst,verbatim}
\usepackage{amsmath,amsfonts,amssymb,amsgen,amsbsy,amsthm,dsfont,color}
\usepackage{MnSymbol,eucal,mathrsfs}
\usepackage[utf8]{inputenc}
\DeclareMathSymbol{\varkappa}{\mathord}{AMSb}{"7B}

\usepackage{color}
\usepackage[square,numbers]{natbib}
\newcommand{\depth}{d}                              
\newcommand{\sur}[1]{{#1}_\text{s}}                    
\renewcommand{\bot}[1]{{#1}_\text{b}}                  

\newcommand{\ud}{\mathrm{d}}

\newcommand{\ui}{\mathrm{i}}
\newcommand{\upi}{{\pi}}

\newcommand{\half}{{\textstyle{1\over2}}}

\newcommand{\third}{{\textstyle{1\over3}}}

\renewcommand{\Re}{\operatorname{Re}}
\renewcommand{\Im}{\operatorname{Im}}

\theoremstyle{remark} 
\newcommand{\eqdef}{\stackrel{\text{\tiny{def}}}{=}}

\newlength{\intwidth}

\newcommand{\cref}{c_{\textsc{r}}}

\title[Extreme wave measurements]{\bf Extreme water wave profile recovery from pressure measurements 
at the seabed}

\author[D. CLAMOND]{Didier CLAMOND}

\newcommand{\nfont}{\fontshape{n}\selectfont}
\address{({\nfont\textbf{Didier Clamond}}) Universit\'e C\^ote d'Azur, CNRS UMR 7351, LJAD, 
Parc Valrose, F-06108 Nice cedex 2, France.} 
\email{didier.clamond@unice.fr}

\author[D. Henry]{David HENRY}

\address{({\nfont\textbf{David Henry}}) Department of Applied Mathematics, University College Cork, 
Cork, Ireland.} 
\email{d.henry@ucc.ie}

\begin{document}

\maketitle

\begin{abstract}
The aim of this note is to examine the efficacy of a recently developed approach to the recovery of 
nonlinear water waves from pressure measurements at the seabed, by applying it to the celebrated extreme 
Stokes wave.  
\end{abstract}


\section{Introduction}

In this note, we examine the efficacy of a recently developed approach to the recovery of nonlinear water 
waves from pressure measurements, by applying it to the celebrated extreme Stokes wave.  
The reconstruction of the water wave surface profile from bottom pressure measurements is an important 
issue for marine engineering applications, but corresponds to a difficult mathematical problem. From the 
viewpoint of the qualitative modelling of water waves, the pressure plays an important role in establishing 
various properties of travelling waves \citep{CS2010,hen11,lyons16}. Measuring the surface of water waves 
directly is extremely difficult and costly, particularly in the ocean, so a commonly employed 
alternative is to calculate the free-surface profile of water waves by way of the so-called pressure transfer 
function, which recovers the free-surface elevation using measurements from submerged pressure transducers. 
The key to the success of this approach is in the derivation of a suitable candidate for the pressure 
transfer function, an issue which is the subject of a large body of experimental and theoretical research 
\citep{BD,TH}. Until recent years, most studies were based on a linear transfer function \citep{ES}, primarily 
due to the intractability of the nonlinear governing equations, and even then for irrotational waves 
(cf. \cite{HT} for linear, and weakly nonlinear, recovery formulae for waves with vorticity).  However, 
the inadequacy of linear formulae for waves of even moderate amplitude is well known  \citep{BD,TH}, 
hence the importance of taking into account nonlinear effects.

In \citep{Cp,D2}, various nonlinear nonlocal equations relating the dynamic pressure on the bed and the 
wave profile were obtained without approximation from the governing equations using different approaches, 
however these formulae are quite entangled and not very transparent. This is a significant impediment in 
the pursuit of practical applications, and alternative approaches have focussed on the reconstruction of 
nonlinear water waves in approximate intermediate and shallow water regimes \citep{BL}. Recently, new 
exact tractable relations were derived which allow a straightforward numerical procedure for deriving the 
free surface from the pressure at the bed \citep{CC}. This surface reconstruction procedure requires the 
numerical resolution of an ordinary differential equation, and in \citep{clam13} it was shown that it is 
possible to solve this ODE analytically by recasting it in terms of an implicit functional equation. The 
approach of \citep{clam13,CC} is quite unique in that it works directly with nonlinear waves in the physical 
plane, as opposed to transforming to a conformal plane, and instead obtains recovery formulae through the 
manipulation of holomorphic mappings of the physical variables. The directness of this approach offers a 
number of advantages, and we demonstrate the efficacy and immediate applicability of this approach, both 
from a theoretical and practical viewpoint, by applying it to the wave-profile recovery problem for Stokes 
extreme wave, or wave of greatest height.

The wave of greatest height was postulated to exist by \citet{sto1880}, who used formal 
arguments to obtain it as the limit of smooth periodic travelling waves. Remarkably, the existence of the 
extreme wave solution with a stagnation point at the crest, along with many important qualitative features (such as convexity, the existence 
of an angular crest with inner angle $2\pi/3$) were rigorously proven only in recent decades (cf. 
\citep{BT,con2016,tol} for surveys of the rigorous analysis of Stokes waves). Also, physical properties 
of the extreme Stokes wave are not obviously transparent: although it is the Stokes wave  of greatest height, 
it is not the most energetic, or fastest, Stokes wave \citep{SF}. While the rigorous mathematical analysis 
of nonlinear Stokes waves is already technically challenging for regular (smooth) waves, this is particularly 
so for the Stokes wave of greatest height, which is  impervious to standard analytical methods due to the 
presence of a stagnation point at the wave-crest. A commonly employed approach for the extreme Stokes wave 
is to use continuation arguments which assume that the extreme wave is a uniform limit of smooth (`almost 
extreme') Stokes waves \citep{BT,const12,lyons16,tol}. Due to the direct nature of the pressure surface 
recovery approach developed in \citep{clam13}, we do not need to  employ convoluted continuation arguments 
in our analysis. This illustrates the robustness and applicability of the reconstruction method of 
\citep{clam13}, which does not invoke intermediate conformal changes of variables, and which accordingly 
offers potential for further development towards capturing more physically complex scenarios, such as the 
recovery of nonlinear rotational water waves. {{For extreme and near-extreme waves, the advantage of working 
in the physical plane, instead of the conformal one, is outlined in \citep{Clamond2018}; the 
present paper provides further support to this claim.}}

\section{Hypothesis and notations}\label{secnot}

We consider two-dimensional periodic travelling waves propagating on the surface of an incompressible 
inviscid fluid under the restoring force of gravity. The wave is periodic\footnote{
Obviously, aperiodic waves are obtained letting $k\to0$. Thus, discussing periodic waves (for 
simplicity) is not a limitation for our purpose here.} in the $x$-direction, with 
period $L$, and the wavenumber is given by $k=2\pi/L$. If $(X,Y)$ denotes horizontal and vertical coordinates 
in a fixed reference frame, with $(U,V)$ the respective velocity field, then the travelling wave hypothesis 
implies a functional dependence on the variables of the form $(X-ct,Y)$: the Galilean change of variables 
\[
x=X-ct, \quad y=Y, \quad u=U-c, \quad v=V,
\]
transforms to a reference frame where the fluid motion is steady, with the resulting velocity field in 
the frame moving with the wave given by $(u,v)$. The wave phase velocity $c$ is 
an arbitrary constant as long as we have not defined what we mean by ``fixed frame of reference''. 
If it is the frame where the horizontal velocity has zero mean at bottom (which corresponds to Stokes' first definition 
of the wave speed), the wave phase velocity $c$ is such that $\left<\/U_\text{b}\/\right>=0$ in the fixed 
frame, that is $\left<\/u_\text{b}\/\right>=-c$ in the frame of reference moving with the wave. Here $\left<\boldsymbol{\cdot}
\right>$ is the Eulerian average operator over one wavelength and, in what 
follows, the subscripts  `$\text{s}$' and `$\text{b}$' denote the restrictions to the free surface and 
at the bed, respectively. Note that $c>0$ if the wave travels toward the increasing $x$-direction; a 
choice that can always be made without lost of generality. Many other definitions of $c$ can of course 
be made, with resulting consequences for the Bernoulli constant, and the gauge condition for the velocity 
potential,  for instance  (cf.  \citep{Clamond2017} for discussions on this, and related, matters).

The horizontal bottom is located at  $y=-\depth$, and the  free surface is described by $y=\eta(x)$: since 
these boundaries are impermeable and steady we have $\sur{v}\/=\/\eta_x\/\sur{u}$ (with $\eta_x\/=\/\ud
\eta/\ud x$) and $\bot{v}\/=\/0$. The mean water level is located at $y=0$, which implies that 
$\left<\eta\right>=0$ for the  ($2\pi/k$)-periodic wave profile $\eta$, that is
\begin{align} \label{defmean}
\left<\,\eta\,\right>\, \eqdef\ \frac{k}{2\,\pi}\int_{-{\pi/k}}^{\,{\pi/k}}\eta(x)\ \ud\/x\ =\ 0.
\end{align}
The local spin or rotation of an infinitesimal fluid element is measured by the vorticity, which is 
expressed by $\omega=u_y-v_x$ for two-dimensional motion. An inherent property of inviscid flows is 
that the vorticity of a given fluid element is preserved by the resultant fluid motion \citep{con2011}, 
and accordingly a fluid mass which is initially irrotational will remain so for all further times. 
Therefore assuming irrotational motion, $u_y=v_x$, allied to the incompressibility condition $\nabla 
\cdot (u,v)=0$, enables us to define a velocity  potential, $\phi(x,y)$,  and stream function, $\psi
(x,y)$, respectively, such that 
$u\/=\/\phi_x\/=\/\psi_y$ and $v\/=\/\phi_y\/=\/-\/\psi_x$.
It follows that the complex potential, defined by $f(z)\/=\/\phi(x,y)\/+\/\ui\/\psi(x,y)$, and the 
complex velocity, $w\/=\ud f/\ud z\/=\/u\/-\/\ui\/v$, are holomorphic functions of the complex abscissa  
$z=x+\ui\/y$ (where $\ui^2=-1$). The stream function $\psi\/=\/\Im\{f\}$ is constant at $y=-\depth$ and 
at $y=\eta$ since these boundaries are impermeable and steady, hence $f(z)$ represents a conformal 
mapping from the fluid domain to a fixed rectangle, the so-called conformal plane, once the non-stagnation 
condition
\begin{equation}\label{stag}
u(x,y) \leqslant \epsilon<0
\end{equation}
holds throughout the fluid domain. 
The  equation of motion is given by the Euler equation which,  for irrotational fluid, can be expressed 
in terms of the following Bernoulli condition 
\begin{align} \label{bernbase}
2\,p\ +\ 2\,g\,y\ + \ u^2\ +\ v^2\ =\ B, \qquad\  
x \in \mathds{R},\ -\depth \leqslant y \leqslant \eta,
\end{align}
where $B$ is the Bernoulli constant, and  $p=p(x,y)$ is the pressure (relative to the atmospheric 
pressure) divided by the (constant) density. At the free surface, surface tension is neglected and 
the relative normalised pressure is zero, $\sur{p}\/=\/0$. From (\ref{defmean}) and (\ref{bernbase}), 
we get
\begin{equation}\label{mt}
B\ =\,\left<\,u_\text{s}^{\,2}\,+\,v_\text{s}^{\,2}\,\right>\,=\,\left<\,u_\text{b}^{\,2}
\, \right>\,>\ 0,
\end{equation}
where the second equality in (\ref{mt}) follows from an application of the divergence theorem applied 
to the vector field $(-2uv,u^2-v^2)$, cf. \citep{CC}.  Note that $B=c^2$ in deep water ($\depth\to\infty$), 
and for solitary waves ($k\to0$), since in these settings $(u,v)\to (-c,0)$ as $y\to -\infty$, and 
$|x|\rightarrow \infty$, respectively. 
Finally we observe that relations (\ref{bernbase}) and (\ref{mt}) yield
\begin{equation} \label{defpmean}
\left<\,\mathfrak{p}\,\right>\,=\ g\,\depth,
\end{equation}
where $\mathfrak{p}(x)\/\eqdef\/p(x,-\depth)=\bot{p}(x)$ is the normalised relative pressure 
at the bed. This is a useful formula for determining the mean fluid depth $\depth$ from pressure 
measurements at the flat bed.

\section{Stokes extreme wave}

In this paper, by `Stokes wave' we mean a periodic, irrotational water wave, with a strictly monotone 
symmetric free-surface profile --- increasing from trough-to-crest and decreasing from crest-to-trough 
--- which moves with a constant speed and with a constant form over a body of water on a flat bed. For 
such waves, the functions $u$, $\eta$, $p$ are even functions and $v$ is an odd function with respect 
to the $x$-variable; we assume the wave crest is located at $x=0$, and the wave troughs are located at 
$x=\pm\pi/k$. It was shown by \citet{sto1880}, using formal methods, that regular wave solutions exist 
for which the strict inequality \eqref{stag} holds throughout the fluid domain. Given this inequality, 
it can be rigorously proven that a continuously differentiable free-surface must {\em a priori\/} be a 
symmetric, real-analytic graph of a function: surveys of  rigorous analytical results for Stokes waves 
can be found in \citep{BT,con2011,con2016,tol}.

Stokes also posited, using heuristic arguments, that a limiting wave solution exists for which \eqref{stag} 
fails and the equality $u=0$ is attained precisely at the wave crest, with the associated breakdown in 
regularity of the wave-profile at the crest manifested in the form of an angular point\footnote{
{In the conformal plane, the curve $\eta(\phi)$ presents a cusp at the crest where 
$\eta\propto|\phi|^{2/3}$ locally, while in the physical plane the curve $\eta(x)$ presents an angular 
point where $\eta\propto|x|$ locally.}} which has a containing angle of $2\pi/3$.   The existence of Stokes 
waves is rigorously established by way of highly-technical bifurcation theory methods and,  indeed, 
it was only in recent decades that a number of fundamental questions concerning the existence of an 
extreme Stokes wave, and properties of the associated wave profile, were rigorously settled (cf. 
\citep{BT,con2016,tol}). For this limiting wave `of greatest height', or extreme Stokes wave, it can be 
shown that the free-surface is real-analytic everywhere except for the stagnation point at the crest, 
where it is merely continuous. We note that the wave crest is an `apparent' stagnation point, with  
fluid particles only coming to rest  there instantaneously, cf. \cite{const12}.

All technical complications arising in the rigorous analysis of extreme Stokes waves essentially stem from 
the lack of regularity at the stagnation point, with the primary impediment being a breakdown in the 
conformality of the mapping $f(z)$ due to the failure of condition \eqref{stag} at the wave-crest. 
Nevertheless,  a detailed asymptotic description of the wave profile, and velocity field, in the 
neighbourhood of the wave-crest is possible. This work originated with \citet{sto1880}, who postulated 
that the wave of maximum amplitude has an included crest angle of $2\pi/3$, the crest being a stagnation 
point which,  in the conformal plane, has a power $2/3$ singularity. It was shown by \citet{grant73} 
that higher-order expansions in the vicinity of the crest involve non-algebraic powers (see also 
\citep{nor74}); this work was placed on a rigorous footing in \citep{AF,mc87}. 
Performing the same local analysis (around the crest at $x = 0$ where $\eta= a$) in the physical plane, 
one obtains after some algebra 
\begin{align}
\eta\, &=\ a\ -\ 3^{-1/2}\,|\/x\/|\ +\ \beta\,\kappa^{\nu-1}\,|\/x\/|^{\nu}\ +\ O\!\left(|x|^{2\nu-1}\right),
\label{etamaxloc}\\
w^2\, &=\ g\,(\/a\/+\/\ui\/z\/)\ +\ \beta\,\kappa^{\nu-1}\left[\,3^{1+\nu}\,4^{-\nu}\,
(1+\nu+\nu^2)\,\right]^{1\over2}g\,(\/a\/+\/\ui\/z\/)^\nu\ +\ O\!\left((a\/+\/\ui\/z)^{2\nu-1}\right),
\label{w2maxloc}
\end{align}
where $\beta$  is a strictly positive (dimensionless) constant \citep{mc87}, $\kappa$ is a (freely 
choosable) characteristic wavenumber (e.g. $\kappa=k$ or $\kappa=g/B$) and $\nu\approx2.20401861065003478$
is the smallest positive root\footnote{The next positive root of \eqref{traneq} 
is $\nu_2\approx5.36>2\nu-1$. There are an infinite number of real positive roots $\nu_j$ 
of \eqref{traneq} such that $\nu_j\sim3j-\half-\left.\sqrt{3}\right/2\pi j$ as $j\to\infty$. There 
are also two complex roots, approximately $-\half\pm\ui\times1.0714$, that do not 
appear in the extreme wave (as they would yield an unbounded free surface); they do play a role 
in near extreme waves, however \citep{LHF77}.} of the transcendental equation \citep{grant73}  
\begin{equation}\label{traneq}
\sqrt{3}\,\tan\!\left(\third\nu\upi\right)\,=\ -\/1\ -\ 2\,\nu^{-1}.
\end{equation}
In order to derive expression \eqref{w2maxloc}, we use (cf. eq. (3.3) in \citep{CC}) the relation 
\begin{equation}\label{wCC}
w_\text{s}^{\,2}\ =\ 2\,g\,(a-\eta)\,(1-\ui\eta_x)\,/\,(1+\ui\eta_x),
\end{equation} and taking  $\eta$ of the form  \eqref{etamaxloc}, 
for some $\nu>1$, equation \eqref{wCC} yields 
\begin{equation}
\frac{w_\text{s}^{\,2}}{g}\ =\ \frac{|\/x\/|}{\sqrt{3}}\ +\ \ui\,x\ +\ \beta\,|\/\kappa\/x\/|^{\nu-1}
\frac{(3\nu-2)\,|x|\,-\,\ui\,(\nu+2)\,\sqrt{3}\,x}{2}\ +\ O\!\left(|x|^{2\nu-1}\right),\label{w2smaxloc}
\end{equation}
that can be conveniently rewritten
\begin{equation}
\frac{w_\text{s}^{\,2}}{g\,(a+\ui\/z_\text{s})}\ =\ 1\ -\ 3\,\beta\,|\/\kappa\/x\/|^{\nu-1}
\frac{\sqrt{3}\,+\,\ui\,(2\nu+1)\,\operatorname{sgn}(x)}{4}\ +\ O\!\left(|x|^{2\nu-2}\right).\label{w2smaxlocbis}
\end{equation}
Then, seeking a complex velocity of the form $w^2=g(a+\ui z)+A(a+\ui z)^\nu+O\!\left((a+
\ui z)^{2\nu-1}\right)$, comparison with \eqref{w2smaxlocbis} leads to $A$ 
as defined by \eqref{w2maxloc} and the relation \eqref{traneq}.
Note the simplicity of this approximation, 
in particular the leading approximation for $w^2\approx g\/(\/a\/+\/\ui\/z\/)$,  which yields 
at once (from the Bernoulli equation) the following approximation for the pressure near the wave crest:
\begin{equation}\label{prescrest}
p(x,y)\ \approx\ g\,(a-y)\ -\ \half\,g\,\sqrt{\,x^2\,+\,(a-y)^2\,}.
\end{equation}
This relation will play an important role in section \ref{secew} below. To the best of our knowledge, 
relations \eqref{w2maxloc} and \eqref{prescrest} have not appeared before in the literature.
A study of the qualitative behaviour of the velocity field for the extreme Stokes near the wave crest 
has been explored in \cite{const12}, where it is shown that the velocity field exhibits the limiting 
behaviours
\refstepcounter{equation}
\[
\lim_{x\to 0} \frac{u^2(x,\eta(x))}{g\,|x|}\,=\,\frac{\sqrt3}{2}, \qquad \lim_{x\to 0} 
\frac{v^2(x,\eta(x))}{g\,|x|}\,=\,\frac{1}{2\,\sqrt3 },
\eqno{(\theequation{\mathit{a},\mathit{b}})}\label{limvel} 
\]
which accord with the limits given by \eqref{w2maxloc} and \eqref{w2smaxloc}.

On the crest line $x=0$ (i.e., for $\Re\{z\}=0$ and $\Im\{z\}\leqslant a$) 
we have $\Re\{w^2\}=u^2\geqslant0$ and $\Im\{w^2\}=0$, the relation \eqref{w2maxloc} then implies 
at once that $\beta$ is necessarily positive. This result was first obtained by \citet{mc87} with 
much more convoluted arguments. However, \citet{mc87} also proved that $\beta\neq0$, the possibility 
$\beta=0$ being not easily ruled out here (that would require more than just a local analysis around 
the crest). 
From the inequality $\beta>0$, the semi-derivatives at the crest are given by
\refstepcounter{equation}
\[
\lim_{x\to0^{\pm}} \eta_{x}\,=\,\mp\,3^{-1/2}, \qquad
\lim_{x\to0^{\pm}} \eta_{xx}\,=\,0^{+}, \qquad 
\lim_{x\to0^{\pm}} \eta_{xxx}\,=\,\pm\,\infty,
\eqno{(\theequation{\mathit{a},\mathit{b},\mathit{c}})}\label{limeta} 
\]
in comparison with the corresponding semi-derivatives in the conformal plane (where, for example, 
$\eta_\phi$ is infinite at the crest since $\eta\propto|\phi|^{2/3}$ locally).

\section{Wave surface recovery: regular waves}\label{secCC}

In this section, we briefly recall the salient features of the surface profile recovery procedures 
presented in \citep{clam13,CC}. The function $\mathfrak{P}$ defined by
\begin{equation}\label{P}
{\mathfrak P}(z)\ \eqdef\ \half\,B\ +\ g\,\depth\ -\ \half\,w^2(z)\ =\ \half\,B\ 
+\ g\,\depth\ -\ \half\,(u^2-v^2)\ +\ \ui\,u\,v,
\end{equation}
is holomorphic, and its restriction at the flat bed coincides with the real-valued function 
$\mathfrak{p}(x)$, since 
\(
\bot{\mathfrak{P}}\/=\/\mathfrak{P}(x-\ui\depth)\/=\/g\/d\/+\/\half\/(\/B\/-\/u_\text{b}^{\,2}\/)
\/=\/\mathfrak{p}(x).
\)
Accordingly ${\mathfrak P}(z)\/=\/{\mathfrak p}(z+\ui\depth)$ and, since $p$ is not a harmonic function 
\citep{con2011,CS2010,hen11}, the flat bed is the only location where $\mathfrak p$ coincides with 
$ \operatorname{Re}\mathfrak{P}$. It can be shown (cf. \citep{clam13,CC}) that
\begin{equation} \label{cde}
g\,\eta\,(1\/-\/\ui\/\eta_x)\ +\ \ui\,B\,\eta_x\ =\ (\,\sur{\mathfrak{P}}\,
-\, g\,\depth\,)\,(1\/+\/\ui\/\eta_x), 
\end{equation}
with $\sur{\mathfrak P}\/=\/\mathfrak{P}(\sur{z})\/=\/\mathfrak{P}(x+\ui\eta(x))$.
The real and imaginary parts of (\ref{cde}) give the two equations 
\begin{align}
g\,\eta\ &=\ \operatorname{Re}\{\mathfrak{P}_\text{s}\}\ -\ g\,\depth\ -\ \eta_x\,
\operatorname{Im}\{\mathfrak{P}_\text{s}\}\ =\ \half\,(\,B\,-\,u_\text{s}^{\,2}\,
-\,v_\text{s}^{\,2}\,), \label{e1} \\
(B\, -\, g\,\eta)\,\eta_x\ &=\ (\,\operatorname{Re}\{\mathfrak{P}_\text{s}\}\,
-\,g\,\depth\,)\,\eta_x\ +\ \operatorname{Im}\{\mathfrak{P}_\text{s}\}\ =\ 
\half\,(\,B\,+\,u_\text{s}^{\,2}\,+\,v_\text{s}^{\,2}\,)\,\eta_x.\label{e2}
\end{align}
where the second equalities are obtained using (\ref{P}) and $\sur{v}\/=\/\eta_x\/\sur{u}$.
From the viewpoint of determining the wave amplitude $a$, for regular Stokes waves we have 
$\eta_x(0)=0$ at the crest $z=\ui\/\eta(0)=\ui\/a$, and so \eqref{e1} reduces to a simple 
implicit equation which can be iteratively solved to find the amplitude $a$ 
as solution of the equation  
\begin{equation}\label{defamp}
a\ =\ g^{-1}\,\Re\{\mathfrak{P}(\ui a)\}\ -\ d.
\end{equation}
The convergence of fixed point iterations of \eqref{defamp} was studied in \citep{CC} 
considering the functional iterations (convergent for all regular waves)
\begin{equation}\label{defiteramp}
y\, =\, \hat{F}(y)\, \eqdef\, y\, +\, g^{-1}\,p(0,y)\, 
=\, g^{-1}\,\Re\{\mathfrak{P}(\ui y)\}\, -\, d\, 
=\, \half\,B\,g^{-1}\, -\, \half\,g^{-1}\,\Re\!\left\{w(\ui y)^2\right\}.
\end{equation}
With the wave amplitude $a$ so determined, in \citep{CC} it was shown that $\eta$ can 
subsequently be obtained by re-expressing (\ref{e2}) as the ordinary differential equation 
\begin{equation}\label{r2}
\eta_x\ =\ \operatorname{Im}\{\mathfrak{P}_\text{s}\}\ /\ \left(\,B\, -\, g\,\eta\, 
-\, \operatorname{Re}\{\mathfrak{P}_\text{s}\}\ +\ g\,\depth\,\right),
\end{equation}
with initial data $\eta(0)=a$, which can iteratively be solved. In \cite{CC} it was rigorously proven 
that these recovery procedures converge.

An alternative, more direct approach to the wave profile recovery problem was instigated in 
\citep{clam13} by considering the holomorphic function $\mathfrak{Q}$ defined by
\begin{align} \label{defQfun}
\mathfrak{Q}(z)\ \eqdef\ \int_{z_0}^z\left[\,\mathfrak{P}(z')\,-\,g\,\depth\,\right]\ud\/z'
\ =\ \int_{z_0}^z\half\left[\,B\,-\,w(z')^2\,\right] \ud\/z', 
\end{align}
where $z_0$ is an arbitrary constant. Taking, as in \citep{clam13}, $z_0=\ui a$ at the origin 
of the free surface, and integrating along the surface, leads to the expression
\begin{equation} \label{defQfunsurfeta}
{\mathfrak{Q}}_\text{s}(x)\ =\ \int_0^xg\,\eta(x')\,\ud\/x'\ +\ 
\ui\left[\,\eta(x)\,-\,a\,\right]\left[\,B\,-\,\half\,g\,a\,-\,\half\,g\,\eta(x)\,\right]. 
\end{equation}
The imaginary part of (\ref{defQfunsurfeta}) yields an implicit relation defining $\eta$ of the form
\begin{equation}\label{reletaimp}
\eta\ =\ g^{-1}\left[\,B\, -\, \sqrt{\,(\/B\/-\/g\/a\/)^2\ -\ 2\,g\,
\operatorname{Im}{\mathfrak{Q}}_\text{s}\}\,}\,\right],
\end{equation}
which is a local algebraic (i.e., neither differential nor integral) equation for the free 
surface $\eta$, and it is easily shown that (\ref{reletaimp}) is an exact implicit solution 
of (\ref{r2}). An explicit expression for $\eta$ can be obtained numerically via fixed point 
iterations, whose convergence is determined as follows. Let be the functional iterations
\begin{equation}\label{defitF}
y\ =\ F(y)\ \eqdef\ g^{-1}\left[\,B\, -\, \sqrt{\,(\/B\/-\/g\/a\/)^2\, -\, 2\,g\,
\operatorname{Im}\{{\mathfrak{Q}}(x+\ui y)\}\,}\,\right], \qquad
y\in[-\depth,\eta(x)],
\end{equation}
for a known holomorphic function ${\mathfrak{Q}}$ at a fixed abscissa $x$ and for a given $B$. 
Iterations of (\ref{defitF}) converge if we can show that $F$ is a contraction, that is, if 
$-1<F_y<1$. The definition of $F$ gives
\begin{align}\label{Fy}
F_y\, =\, \frac{\left[\,\operatorname{Im}\{{\mathfrak{Q}}\}\,\right]_y}
{\sqrt{\,(\/B\/-\/g\/a\/)^2\, -\, 2\,g\,\operatorname{Im}\{{\mathfrak{Q}}\}\/}}\, 
=\, \frac{p\, +\, g\,y\, +\, v^2}{B\, -\, g\,y}\, =\, 
\frac{p\, +\, g\,y\, +\, v^2}{2\,p\,+\,g\,y\,+\,u^2\,+\,v^2},
\end{align}
where we have used definitions (\ref{P}), (\ref{defQfun})  and the Bernoulli equation (\ref{bernbase}).
The lower bound $F_y>-1$ yields the inequality $p+v^2+B>0$ that is obviously always satisfied.
The upper bound for contraction condition $F_y<1$ yields the inequality $p+u^2>0$ 
that is always satisfied when \eqref{stag} holds. Therefore, for regular Stokes waves, $F$ is a 
contraction everywhere in the bulk and at the free surface and the 
iterations (\ref{defitF}) converge to an unique solution.

Note that, since $F_y$ is independent of $a$ (and so independent of the choice of  $z_0$ in the 
definition of $\mathfrak{Q}$), the convergence proof for the iterations (\ref{defitF}) is valid 
wherever $z_0$ is chosen in the fluid.
Furthermore, the analysis above for the wave profile recovery 
procedure is valid for all regular waves, periodic or not, with one or more crests and troughs per 
period, symmetric or not. It is not contingent on the waves being Stokes waves.

\section{Wave surface recovery: extreme waves}\label{secew}

The above considerations may breakdown for the extreme waves, where \eqref{stag} 
fails to hold at the stagnation point at a wavecrest. In this case the wavecrest located at 
$(0,\eta(0))=(0,a)$ takes an angular form with containing angle $2\pi/3$, with an asymptotic description 
provided by \eqref{etamaxloc}, and so in particular $\eta_x(0^\pm)\/\neq\/0$. The analysis of the 
previous section (see also \citep{clam13,CC}) then needs to be refined.

\subsection{Crest determination}

The analysis of the functional iterations (\ref{defitF}) assumes that the constants 
$a$ and $B$ are known. This is fine for a mathematical proof, but not so for practical applications 
because $a$ and $B$ are generally unknown {\em a priori}. Since the crest of an extreme wave is a 
stagnation point where $u=v=p=0$, it follows directly from \eqref{bernbase} that $B\/=\/2\/g\/a$, 
so only $a$ remains to be determined before studying the iterations \eqref{defitF}.  
The convergence of fixed-point iterations of \eqref{defamp}  determining the wave amplitude for extreme 
waves, where $\eta_x\neq0$ at the crest, was not addressed in \citep{CC}, however we remark that formula 
\eqref{defamp} is still applicable in this regime since $ \operatorname{Im}\{\mathfrak{P}_\text{s}\}=0$ 
along the crest-line (where $v=v_y=0$):  fixed-point iterations determined by relation \eqref{defiteramp} 
converge if $-1<\hat{F}_y<1$. At the crest of an  extreme wave, employing the asymptotic relation 
\eqref{prescrest} (which is valid in a neighbourhood of the crest) we can compute $\hat{F}_y=1+g^{-1}
p_y(0,y)=1/2$, so it follows that the fixed point iterations defined by \eqref{defiteramp} converge. \\

\subsection{Remarks}

{\it i}- On the crest line of a Stokes wave the inequality $-g\leqslant p_y<0$ holds: although the 
second inequality 
is physically quite intuitive, the rigorous proof of this fact for nonlinear waves relies on a fine technical 
analysis of harmonic functions for regular waves \citep{CS2010,hen11}, with further complications needing to 
be addressed in the extreme wave setting \cite{lyons16}. These bounds are sharp with respect to the analytical 
methods being employed. In this context, it is interesting that the asymptotic approximation \eqref{prescrest} 
yields that $p_y=-g/2<0$ at the crest of an extreme wave:  one might be tempted to conjecture as to whether 
the inequality  $-g\leqslant p_y\leqslant-g/2$ holds on the crest line of extreme waves and, perhaps, of 
regular Stokes waves which converge to the extreme wave by the standard limiting process \cite{const12,tol}.  
As a caveat to the latter postulation, we note that many physical properties  of the extreme Stokes wave are not 
obviously transparent; for instance, although it is the Stokes wave  of greatest height, it is not the most 
energetic, fastest, or impulsive of the Stokes waves \citep{SF}.

{\it ii}- A further consequence of the relation $p_y=-g/2$ at the crest of an extreme wave is that, since 
the relation $u\partial_x\left[ v(x,\eta(x)) \right]=-p_y-g$  holds along the wave profile $y=\eta(x)$ for 
all Stokes waves, we must have $\partial_x\left[ v(x,\eta(x)) \right]$ blowing up as $x\to 0$ since $u$ 
vanishes at the wave crest. It follows that $u_y=v_x$ must also blow-up the wave-crest: this can also be 
seen directly from setting $-p_y-g=-g/2$ in the second component of the Euler equation.  

{\it iii}- A useful relation involving the wave profile, particularly in the context of Stokes extreme wave, 
is obtained by multiplying (\ref{e2}) by $\eta_x$ and adding (\ref{e1}), from which  one gets (after 
some elementary algebra):
\begin{align}
g\,\eta\  &=\ \operatorname{Re}\{\mathfrak{P}_\text{s}\}\ -\ g\,\depth\ -\ 
(B\, -\,2\,g\,\eta)\,\eta_x^{\,2}\,\left(\,1+\,\eta_x^{\,2}\,\right)^{-1}.\label{e5}
\end{align}
At the crest of an extreme wave we have  $\eta_x\neq0$ is finite, since the angular crest is not a cusp, 
and $B-2g\eta=0$, and so \eqref{e5} reduces to  relation \eqref{defamp} at the crest.

\subsection{Surface determination}

The convergence of the functional iterations (\ref{defitF}) is proven above (see \citep{clam13} 
for more details) everywhere except at an angular crest where $F_y(a)=1$: this corresponds to a 
so-called {\em positively neutral\/} fixed point \citep{bair97}. To determine if the iterations 
converge where $F_y=1$, we can consider the second derivative of $F$: 
\begin{equation}\label{defFyy}
F_{yy}\ =\ \frac{g\,F_y\,+\,p_y\, +\, g\, +\,2\,v\,v_y}{2\,p\,+\,g\,y\,+\,u^2\,+\,v^2}\ 
=\ \frac{g\,F_y\,+\,v\,v_y\,-\,u\,v_x}{B\,-\,g\,y}\ 
=\ \frac{g\,F_y\,+\,\operatorname{Im}\{\/w\,w_z\/\}}{B\,-\,g\,y}.
\end{equation}
In order to estimate $F_{yy}$ at angular crests, we use the expansions \eqref{etamaxloc} and 
\eqref{w2maxloc}, to find that $F_{yy}=3/2a>0$ at the angular crests. Hence, the fixed point 
$y=a$ is monotonously semi-stable from below \citep{bair97}, which  means that the iterations are convergent 
when the fixed point $y=a$ is approached from the fluid side $y<a$, and divergent when approached 
from above the surface  $y>a$.
For practical applications, this results suggests that basic fixed-point iterations of 
\eqref{reletaimp} convergence very slowly in the vicinity of an angular crest. 
More sophisticated iterations (such as Newton and Levenberg--Marquardt methods) should then 
be used instead. It is not our purpose here to investigate these possibilities, however.

\subsection{Trough determination}

At the wave trough, where  $\eta=-b\leqslant0$ and so $a+b$ is the total wave height, for all Stokes waves 
including the extreme wave, we have $\eta_x(\pm \pi/k)=0$. Equation \eqref{e1} then reduces to the 
implicit equation 
\begin{equation} \label{r1t}
b\ =\ \depth\ -\ g^{-1}\,\operatorname{Re}\{\mathfrak{P}(-\ui\/b)\}.
\end{equation}
In \citep{clam13}, presented numerical evidence suggests that iterations of 
(\ref{r1t}) converge to $b$, however a mathematically rigorous proof of this result is still lacking. 
Indeed, as shown in \citep{clam13}, convergence is guaranteed if $-2g<p_y<0$ on the trough line 
$y\in[-d,-b]$. The upper bound is ensured by the inequality $p_y\leqslant-g$ on the trough line, which 
can be rigorously established for nonlinear Stokes waves of all amplitudes  \cite{con06,CS2010,hen11,lyons16} 
by way of maximum principles, using the fact that the pressure distribution is a super-harmonic function 
with respect to the physical variables, cf. \cite{con2011}. However, a rigorous proof of the lower bound 
for $p_y$ is currently lacking, being unachievable by this approach. This is not a problem for extreme 
waves since $b$ can be computed without relying on (\ref{r1t}). Indeed, $a$ being given by \eqref{defamp}, 
the Bernoulli constant is $B=2ga$ and $b$ is obtained directly by computing \eqref{reletaimp} at a trough.

\section{Conclusion}\label{secconclu}

In this note, we examined the efficacy of exact relations for the recovery of nonlinear free surface 
wave profiles from pressure measurements at the bed, derived in \citep{clam13,CC}, in particular 
their applicability to the reconstruction of the extreme Stokes wave. The extreme Stokes wave possesses 
intrinsic mathematical complications even beyond the level exhibited by regular nonlinear waves, due to 
the presence of a stagnation point at the wave-crest. There is a breakdown in regularity at the crest of 
an extreme Stokes wave, whereby the profile is no-longer differentiable, but merely continuous: 
the sharp wave-crest has an inner angle $2\pi/3$. 

The relations introduced in \citep{clam13} employ a relatively simple local formulation without derivatives 
and integrations, which allows for great scope in their practical application. The directness of the 
surface reconstruction approach initiated in \citep{clam13} is due to the fact that physical variables, 
and the physical plane, are used throughout: intermediate conformal transformations are not invoked. 
The holomorphic function $\mathfrak{Q}$ allows a reformulation of the problem into a simplified, but 
still exact, form.

The convergence of the involved fixed-point iterations, required for obtaining the surface explicitly, 
was examined for the case of limiting extreme waves. A local analysis suggests (semi) convergence at 
angular crests, thereby offering an  insight into the applicability of the reconstruction  method in 
such an extreme case. Practical reconstructions of regular waves are demonstrated 
in \citep{clam13,CC} via numerical examples, so they do not need to be reproduced here. An illustration  
for the extreme wave would require data (experimental or numerical) for the pressure at the seabed, 
that are not available. Beside the possible reconstruction of the extreme wave, one of our goals here 
is to illustrate the usefulness of working in the physical plane, since we could gain theoretical 
insights with only elementary mathematical tools, in addition to a practically useful method.
      
It is hoped that the demonstrated robustness of this wave surface reconstruction approach will prove 
useful in treating other water wave problems, for instance the recovery of fully nonlinear rotational 
water wave profiles from pressure measurements, an objective which has hitherto proven unattainable. 
This will be the subject of future investigations.

\section*{Acknowledgements}
D. Clamond acknowledges the support of the Universit\'e C\^ote d'Azur under the research grant 
{\em Paul Montel Chair\/} 2019.
D. Henry acknowledges the support of the Science Foundation Ireland (SFI) under the research 
grant 13/CDA/2117.

\addcontentsline{toc}{section}{References}
\bibliographystyle{abbrvnat}

\end{document}